\documentclass[12pt]{article}
\usepackage{graphicx,color}
\begin{document}
% Zeilenabstand (doppelt=8.2mm , ca. einfach:5.5mm)
\baselineskip=5.5mm
\newcommand{\be} {\begin{equation}}
\newcommand{\ee} {\end{equation}}
\newcommand{\Be} {\begin{eqnarray}}
\newcommand{\Ee} {\end{eqnarray}}
\renewcommand{\thefootnote}{\fnsymbol{footnote}}
\def\a{\alpha}
\def\b{\beta}
\def\g{\gamma}
\def\G{\Gamma}
\def\d{\delta}
\def\D{\Delta}
\def\e{\epsilon}
\def\k{\kappa}
\def\l{\lambda}
\def\L{\Lambda}
\def\t{\tau}
\def\om{\omega}
\def\Om{\Omega}
\def\s{\sigma}
\def\lg{\langle}
\def\rg{\rangle}
\newcommand{\tblue}[1]{\textcolor{blue}{#1}}
\begin{center}
{\large {\bf Memory effects in the relaxation of the Gaussian trap model} }\\
\vspace{0.5cm}
\noindent
{\bf Gregor Diezemann$^*$ and Andreas Heuer$^\dagger$ } \\
{\it
$^*$Institut f\"ur Physikalische Chemie, Universit\"at Mainz,
Jakob-Welder-Weg 11,\\ 55128 Mainz, FRG\\
$^\dagger$Institut f\"ur Physikalische Chemie, Universit\"at M\"unster,
Corrensstr. 30, 48149 M\"unster, FRG}
\end{center}

\vspace{0.75cm}
\noindent
{\it
We investigate the memory effect in a simple model for glassy relaxation, a trap model with a Gaussian density of states. 
In this model thermal equilibrium is reached at all finite temperatures and therefore we can consider jumps from low to high temperatures in addition to the quenches usually considered in aging studies.
We show that the evolution of the energy following the Kovacs-protocol can approximately be expressed as a difference of two monotonously decaying functions and thus show the existence of a so-called Kovacs hump whenever these functions are not single exponentials.
It is well established that the Kovacs effect also occurs in the linear response regime and we show that most of the gross features do not change dramatically when large temperature jumps are considered.
However, there is one distinguishing feature that only exists beyond the linear regime which we discuss in detail.
For the memory experiment with 'inverted' temperatures, i.e. jumping up and then down again, we find a very similar behavior apart from an opposite sign of the hump.
}

\vspace{1cm}
\section*{I. Introduction}
The out-of-equilibrium dynamics of glassy systems has been studied for a long time, 
for a review see\cite{Bouchaud:1998}. 
Many investigations have been devoted to the study of the violations of the fluctuation-dissipation theorem, cf. ref.\cite{Crisanti:2003p3550}. 
In particular, the question regarding a meaningful definition of an effective temperature still remains to be answered as there are conflicting results, for a recent investigation see\cite{JabbariFarouji:2008p4149} and the literature therein.
Also the aging behavior of time-correlation functions and in particular the rotational and translational dynamics has been investigated with the result that the aging properties of glass-forming liquids are determined by the primary relaxation, cf. 
ref.\cite{Thurau:2002p4151, Lunkenheimer:2005p4153}.

Another effect that has been analyzed in much detail both theoretically and experimentally in the past decade is the so-called Kovacs effect or memory effect
\cite{Kovacs:1963,Kovacs:1979p3646}.
In the original version of the experimental protocol one equilibrates a glass-forming liquid (low molecular or polymeric) at a given temperature $T_0$. Afterwards, a temperature jump to a lower temperature $T_1$ is performed, followed by an increase in temperature to a final temperature $T_2$ after a time $t_1$ and the volume of the sample is monitored as a function of time. 
The crucial point in the experiment is that the time $t_1$ for the up-jump $T_1\to T_2$ is chosen such that $V_{T_1}(t_1)$ coincides with the equilibrium volume at $T_2$, 
$V_{T_1}(t_1)=V_{T_2}^{\rm eq}$.
Instead of staying constant at $V_{T_2}^{\rm eq}$, the volume first increases and then decreases again as a function of time and $V_{T_2}^{\rm eq}$ is only reached in the long-time limit.
The observation of this so-called Kovacs hump usually is taken as an indication for the fact that the system cannot be described fully in terms of time-independent thermodynamical quantities. 
The relaxation appears to depend sensitively on the history of the sample.
This memory effect has not only been observed in the evolution of thermodynamic variables but also shows up in the dielectric constant in both, polymer glasses\cite{Bellon:2000} and a molecular liquid\cite{Yardimci:2003}.
Furthermore, its occurence is not restricted to the slow time scale of the primary relaxation but it has also been found in the region of the low-frequency vibrations of a glass\cite{Wypych:2006}.
The Kovacs effect and its interpretation in terms of simple models has been described in a review-like paper by Bertin et al.\cite{Bertin:2003p2817}. 
Among the models investigated is the so-called trap model with an exponential density of states (DOS)\cite{Monthus:1996p3207}. 
Also computer simulations of model glass-forming liquids\cite{Mossa:2004} and the Ising spin 
glas\cite{Berthier:2002p2819} show the Kovacs effect as well as a number of models showing glassy relaxation, such as the p-spin model\cite{Cugliandolo:2004p2938}, facilitated or constrained spin models\cite{Arenzon:2004p2928,Buhot:2003p3051,Aquino:2006p2885} and also the two-level 
model\cite{Aquino:2008}.
The effect has furthermore been observed in other complex systems like granular 
materials\cite{Josserand:2000} and also in a model for critical dynamics\cite{Berthier:2002}.

If one chooses to perform the experiment with electrical fields as a variable instead of the temperature, one can perform so-called crossover experiments in the linear response 
regime and it was shown that the hump has its origin in the fact that one subtracts two relaxation functions which are evaluated at different times\cite{Boehmer:1995p4162}.
This means, one has to build $\d_1\Phi(t_1+t)-\d_2\Phi(t)$ with $\Phi(t)$ denoting the dielectric (step) response function and the $\d_k$ are the relative field intensities.
Thus, in the linear regime, the Kovacs effect should be observable in all systems exhibiting non-exponential relaxation.
Of course, this finding has its origin solely in the superposition principle of linear response theory and here the Kovacs hump is not at all related to any thermal history.
A discussion of the behavior of the Kovacs hump in linear response theory for Markov processes has  been presented recently\cite{Prados:2010p5466}. 

In the present paper we study the Kovacs effect in the trap model with a Gaussian DOS.
In contrast to the model with an exponential DOS, the system can reach equilibrium in the long time limit for all temperatures\cite{DYRE:1995p2812}. 
Additionally, this model has been shown to qualitatively reproduce some relevant features of the energy landscape of simulated glassy systems\cite{Denny:2003p3738,Heuer:2008p3506}.
In particular, the relaxation time for a given energy has been shown to coincide with the escape rate from a meta-basin up to a factor of two.
Furthermore, after a temperature jump the temporal evolution of the distribution of trap energies behaves similar to what has been observed in computer simulation studies of the distribution of inherent structure or meta-basin energies during aging of a model glass-forming liquid\cite{Diezemann:2005p3083, NicChris:2010}. 
After a quench, the width of the distribution decreases and then increases again as a function of the time that has elapsed and for a jump from low to high temperatures one finds a bimodal distribution of meta-basin energies as predicted by the Gaussian trap model\cite{DYRE:1995p2812}.
Below, we will briefly recall the non-exponential relaxation of the energy after both, a quench from a high temperature and an up-jump from a low temperature.
Also for the Kovacs protocol we will in addition to the standard procedure treat the situation in which the system is in equilibrium at low temperatures initially. Then a jump to a higher temperature and afterwards back to some intermediate temperature is performed and the energy is monitored as in the original protocol.
Furthermore, we will consider the Kovacs experiment for very small temperature differences in the linear response regime.
\section*{II. Non-exponential energy relaxation in the Gaussian trap model}
Before we discuss the Kovacs effect in the Gaussian trap model in detail, we briefly review the aging-properties of the model\cite{Diezemann:2007p3776} and the relaxation after a temperature-jump\cite{NicChris:2010}.

\begin{figure}[!h]
\vspace{-0.3cm}
\centering
\includegraphics[width=7cm]{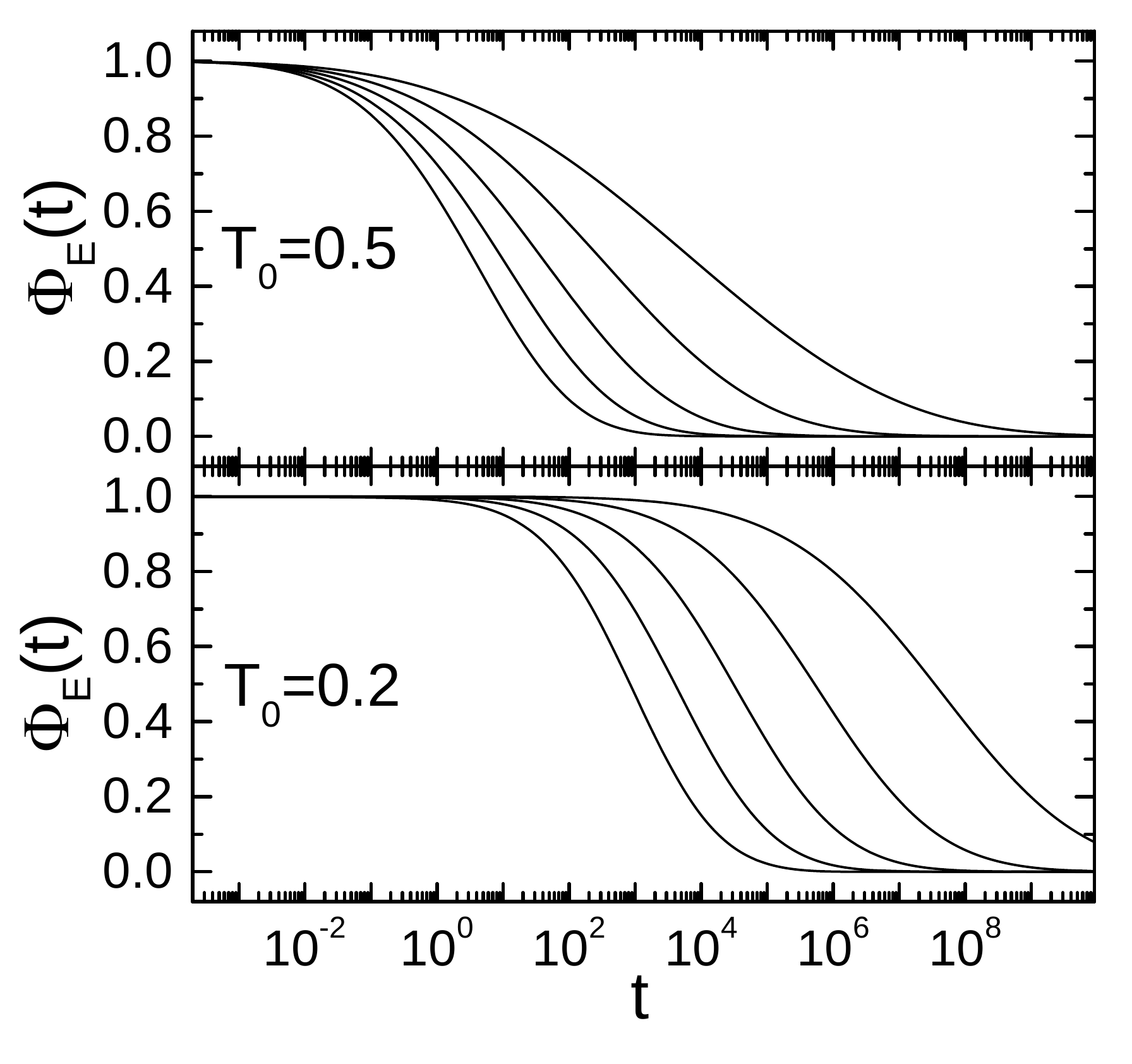}
\vspace{-0.3cm}
\caption{Relaxation function $\Phi_E(t)$ versus time (in units of $\k_\infty$) for different temperatures (in units of $\s$).
Upper panel: jump from $T_0=0.5$ to $T=0.45, 0.4, 0.35, 0.3, 0.25$ from left to right.
Lower panel: jump from $T_0=0.2$ to $T=0.45, 0.4, 0.35, 0.3, 0.25$ from left to right.
}
\label{Fig.One}
\end{figure}

The trap model is defined by the following master equation for the conditional probability (Green's function) of finding the system in a trap with trap energy $\e$ at time $t+t_0$, given it occupied the trap with energy $\e_0$ at time $t=t_0$ ($G_T(\e,t+t_0|\e_0,t_0)=G_T(\e,t|\e_0,0)\equiv G_T(\e,t|\e_0)$ due to time-translational 
invariance)\cite{Monthus:1996p3207, DYRE:1995p2812, NicChris:2010, Diezemann:2007p3776}:
\be\label{ME.G}
{\dot G}_T(\e,t|\e_0)= -\k_T(\e)G_T(\e,t|\e_0)
                         +\rho(\e)\!\int\!d\e'\k_T(\e')G_T(\e',t|\e_0) 
\ee
The subscript $T$ is used to emphasize that the Green's function is defined at temperature 
$T=1/\b$.
In eq.(\ref{ME.G}), the escape rate is given by
\be\label{k.T}
\k_T(\e)=\k_\infty e^{\b\e}
\ee
with the attempt rate $\k_\infty$.
We will measure time in units of $\k_\infty$ in all following calculations.
Furthermore, we solely consider the model with a Gaussian DOS
\be\label{DOS.Gauss}
\rho(\e)\!=\!{1\over\sqrt{2\pi}\s}e^{-\e^2/(2\s^2)}
\ee
with $\s=1$ and throughout the paper we measure temperature in units of $\s$. 
From eq.(\ref{ME.G}), the equilibrium populations at a given temperature $T$ are found to be Gaussian $p^{\rm eq}_T(\e)=\lim_{t\to\infty}G_T(\e,t|\e_0)={1\over\sqrt{2\pi}\s}e^{-(\e-{\bar\e}_T)^2/(2\s^2)}
$
with ${\bar\e}_T=-\b \s^2$.
The solution of eq.(\ref{ME.G}) allows the computions of all quantities of interest.
For instance, the population of the trap with trap energy $\e$ is given by
\be\label{pop.t}
p_T(\e,t)=\int\!d\e_0G_T(\e,t-t_0|\e_0)p(\e_0,t_0)
\ee
where the integral $\int\!d\e$ is to be read as a short hand notation for $\int_{-\infty}^\infty\!d\e$ and $p(\e_0,t_0)$ denotes the population at the initial time $t_0$.
In all calculations performed in the present paper we assume that an instantaneous temperature jump from $T_0$ to $T$ took place immediately before $t_0$. 
This means that the initial population at the destination temperature is given by the equilibrium population at the reference temperature $T_0$, 
\be\label{p.0.eq}
p(\e_0,t_0)=p^{\rm eq}_{T_0}(\e_0).
\ee
Eq.(\ref{pop.t}) allows the computation of general expectation values, for instance moments of the energy-distribution:
\be\label{E.n.t}
E^n_T(t)=\int\!d\e \e^np_T(\e,t).
\ee
The relaxation behavior of the energy has been reviewed briefly in ref.\cite{NicChris:2010}. 
The mean value $E_T(t)$ changes monotonously from the equilibrium value at the initial temperature $T_0$, $E(t=0)=E_{T_0}^{\rm eq}=-\b_0\s^2$ towards the equilibrium value at the destination temperature $T$, $E(\infty)=E_{T}^{\rm eq}=-\b\s^2$.
Also the change of the width of the distribution of the populations of the traps for intermediate times and in particular the intermittant two-peak structure after a temperature up-jump has been discussed.
For our present purpose these latter observations are more important because the change in the width of the populations $p_T(\e,t)$ gives rise to varying degrees of non-exponentiality in the energy relaxation.
This feature becomes apparent when the normalized relaxation function 
\[
\Phi_E(t)=(E_T(t)-E_T(\infty))/(E_T(0)-E_T(\infty))
\]
with $E_T(0)=-\b_0\s^2$ and $E_T(\infty)=-\b\s^2$ is recorded, cf. Fig.\ref{Fig.One}.
One can see that the relaxation time increases with decreasing temperature $T$ independent of the initial temperature $T_0$ and the non-exponentiality of the decay becomes more pronounced.
For the jump from a low temperature, this behavior might be counter-intuitive on first glance. 
However, it just reflects the relaxation behavior at the destination temperature and this relaxation is faster the higher the temperature is.

\begin{figure}[!h]
\vspace{-0.3cm}
\centering
\includegraphics[width=7cm]{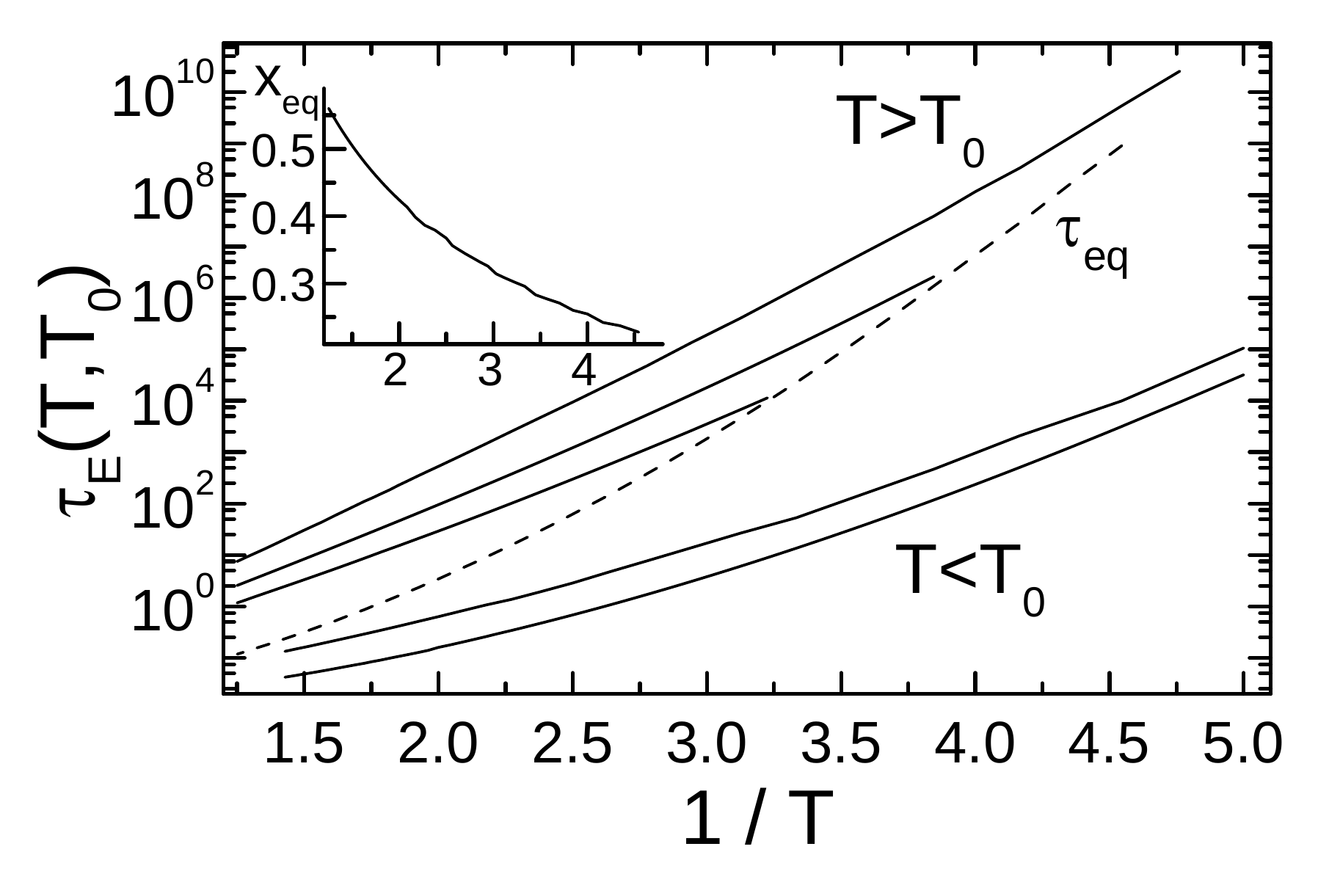}
\vspace{-0.5cm}
\caption{$\t_E(T,T_0)$ for $T_0=\infty$, $T_0=1$, $T_0=0.3$, $T_0=0.25$ and $T_0=0.2$ from bottom to top as a function of the inverse destination temperature $\b=T^{-1}$.
Additionally shown is the equilibrium value $\t_{\rm eq}$ (dashed line) and the stretching parameter $x_{\rm eq}$ (inset) obtained from a fit of $C_{\rm eq}(t)$ to a KWW-function 
($\exp{\left[-(t/\t_{\rm eq})^{x_{\rm eq}}\right]}$).}
\label{Fig.Two}
\end{figure}

The relaxation times $\t_E(T,T_0)$ defined as the $1/e$-decay times of $\Phi_E(t)$ for a variety of initial temperatures $T_0$ as a function of the final temperature $T$ are shown in 
Fig.\ref{Fig.Two}. 
It is apparent that for a given temperature $T$ the value of the relaxation time strongly depends on the initial temperature and that $\t_E$ increases with decreasing $T_0$ as is also observed from 
Fig.\ref{Fig.One}.
This is just the same behavior as is observed if $\t_E$ is considered as a function of $T$.
In addition to the relaxation times of $\Phi_E(t)$ we show the relaxation time $\t_{\rm eq}$ of the normalized energy autocorrelation function $C_{\rm eq}(t)$
\be\label{CN.eq}
C_{\rm eq}(t)={\lg\e(t)\e(0)\rg-(E_T^{\rm eq})^2\over \s^2}
\quad\mbox{with}\quad
\lg\e(t)\e(0)\rg=\int\!d\e\!\int\!d\e_0\e\e_0G_T(\e,t|\e_0)p_T^{\rm eq}(\e_0).
\ee
The relaxation time is obtained from a fit to a KWW-function, 
$C_{\rm eq}(t)\sim\exp{\left[-(t/\t_{\rm eq})^{x_{\rm eq}}\right]}$.
The energy autocorrelation function $C_{\rm eq}(t)$ is of relevance because it determines the energy relaxation in the linear response regime, which is explicitly given by
$E_T(t)=E_T^{\rm eq}+\left(E_{T_0}^{\rm eq}-E_T^{\rm eq}\right)C_{\rm eq}(t)$.
As it is always the lower temperature that determines the value of $\t_E(T,T_0)$ it is not astonishing that $\t_{\rm eq}$ has values similar to the corresponding $\t_E$.
The stretching shows the same qualitative behavior as also observed for other observables in the framework of the Gaussian trap model\cite{Monthus:1996p3207, Diezemann:2007p3776}.

Instead of further discussing the relaxation properties of the Gaussian trap model after fast temperature jumps, we refer the interested reader to 
refs.\cite{DYRE:1995p2812,Diezemann:2005p3083,NicChris:2010} where a detailed discussion of the temporal evolution of the trap-populations $p_T(\e,t)$ can be found.
We close the discussion of the energy relaxation in summarizing that the energy relaxes towards equilibrium in a highly non-exponential way with both, the relaxation time and the stretching strongly depending on temperature. 
These features will be essential for our understanding of the origin of the so-called Kovacs hump.
\section*{III. Kovacs-Effect}
We will now discuss the memory effect for two situations, first starting from a high temperature and afterwards we consider the 'inverse' situation starting from a low temperature. 
In both cases we will consider the usual situation in which the temperature differences are not restricted to be small. 
The relevant questions that we will consider are whether the results for the Gaussian trap model are compatible with the features observed in more complex models and if the two-peak structure observed in the distributions of trap energies after an up-jump has some impact on the results of the 'inverse' Kovacs experiment.
Finally, we will discuss the effect of the experiment in the linear response regime in order to see if there are intrinsic nonlinear effects that give rise to distinctive features for large temperature jumps.
In all cases we choose the energy as the relevant observable.
\subsubsection*{1. 'Standard' Kovacs experiment}
\begin{figure}[!h]
\centering
\includegraphics[width=6cm]{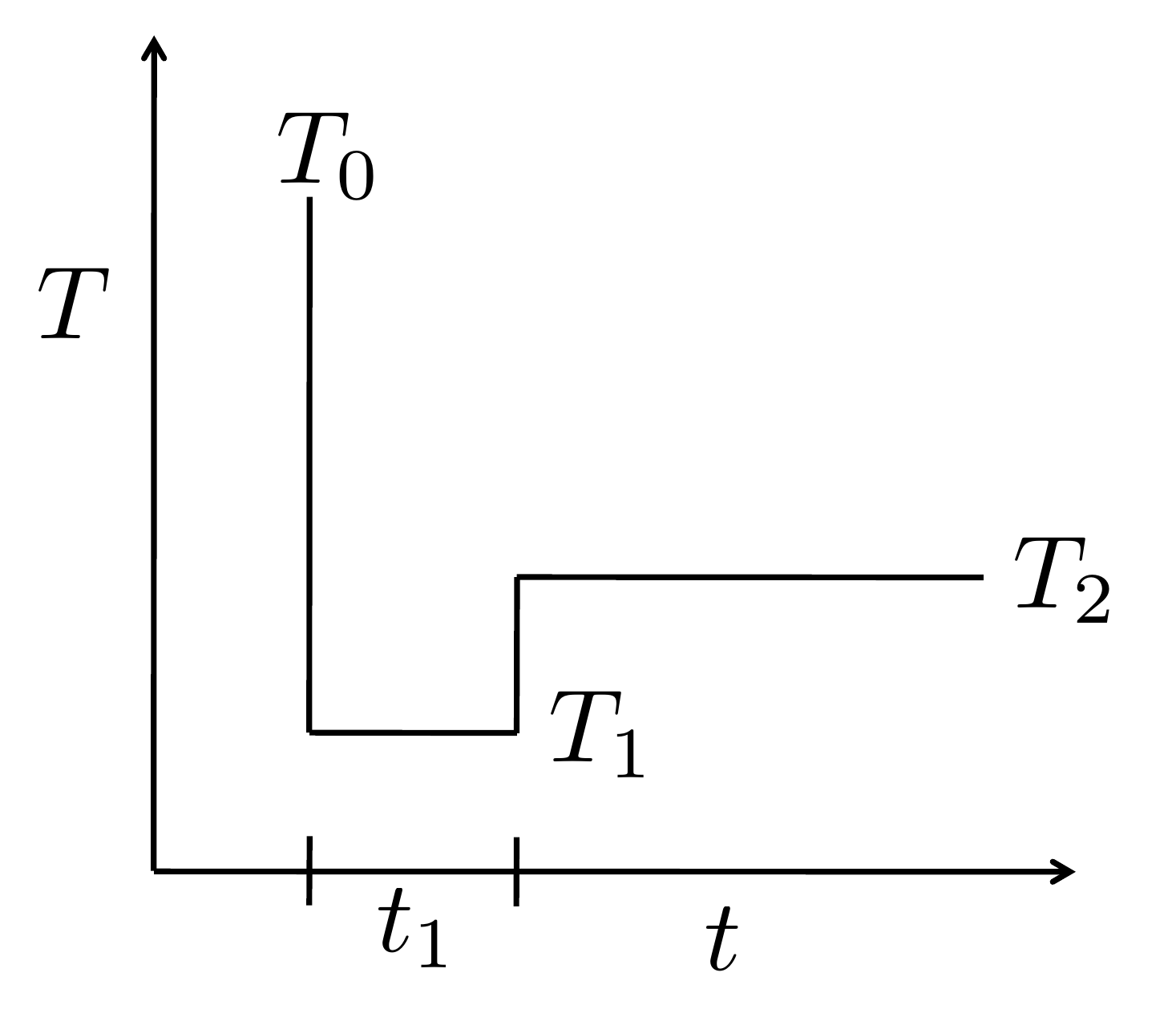}
\vspace{-0.5cm}
\caption{Sketch of the experimental protocol for the Kovacs effect: In the beginning, a quench from $T_0$ to $T_1$ is performed. Then the system is kept at $T_1$ until the energy has reached the equilibrium value at a temperature $T_2$ ($T_1<T_2<T_0$).
At this time $t_1$, the temperature is instantaneously changed to $T_2$ and the temporal evolution of the energy is monitored.}
\label{Fig.Three}
\end{figure}

We start from a high initial temperature and follow the usual protocol, cf. Fig.\ref{Fig.Three}: 
\begin{itemize}
\item
quench:\hspace{2cm} $T_0\to T_1$ ($T_1<T_2<T_0$)
\item 
after a time $t_1$ chosen such that $E_{T_1}(t_1)=E_{T_2}^{\rm eq}\;$:\hspace{0.3cm}
jump to $T_2$.
\item
measurement of the energy $E_{T_2}(t)$ (time $t=0$ at the beginning of the measurement).
\end{itemize}
\vspace{-0.0cm}
From eqns.(\ref{pop.t}) and (\ref{E.n.t}) one finds for the temporal evolution of the energy
\be\label{E.Kovacs}
E_{T_2}(t_1+t)=\int\!\!d\e\!\cdot\!\e\!\cdot\!p_{T_2}(\e,t_1;t)
=\int\!\!d\e\!\int\!\!d\e'\!\cdot\!\e\!\cdot\!G_{T_2}(\e,t|\e')p_{T_1}(\e',t_1)
\ee
where the populations at $T_1$ follow from
$p_{T_1}(\e',t_1)=\int\!\!d\e_0G_{T_1}(\e',t_1|\e_0)p^{\rm eq}_{T_0}(\e_0)$.
(Note that according to our notation defined in eq.(\ref{pop.t}) one has 
$p_{T_2}(\e,t_1;t)\equiv p_{T_2}(\e,t_1+t)$.)

The definition of $t_1$, $E_{T_1}(t_1)=E_{T_2}^{\rm eq}$, makes clear that $E_{T_2}(t_1+t)$ can conveniently be written as 
\be\label{DelE.def}
E_{T_2}(t_1+t)=E_{T_2}^{\rm eq}+\D E(t)
\ee
thus defining the quantity $\D E(t)$, the so-called Kovacs hump.
For this, one has by definition $\D E(0)=0$ and additionally $\D E(t\to\infty)=0$.
For a more detailed discussion of the above definition see ref.\cite{Bertin:2003p2817}.

\begin{figure}[!h]
\vspace{-0.2cm}
\centering
\includegraphics[width=7cm]{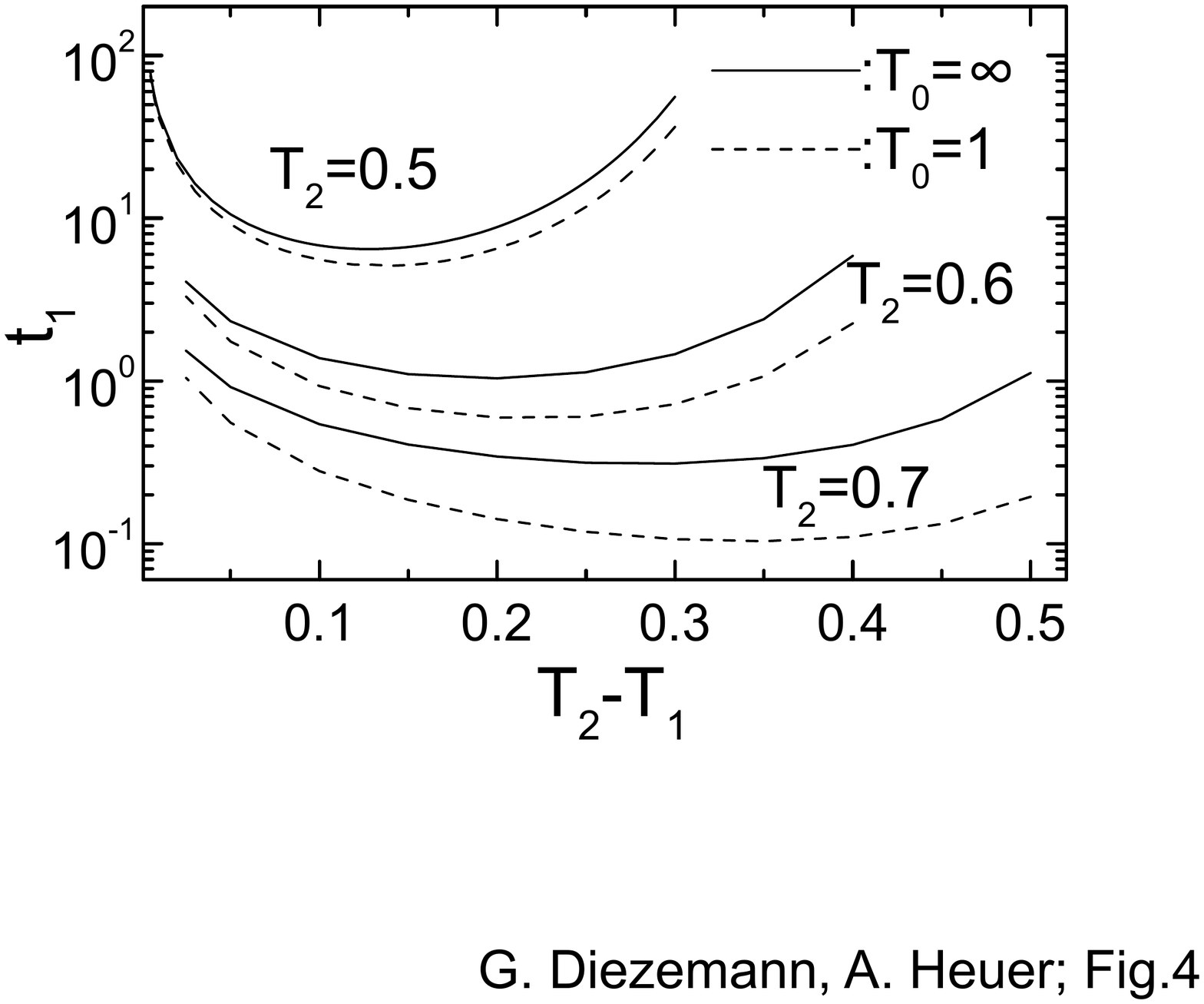}
\vspace{-0.3cm}
\caption{$t_1$ for $T_2=0.5, 0.6, 0.7$ following a quench from $T_0=\infty$ to various 
$T_1\in(0.2\cdots 0.5)$ (full lines) and the same for a quench from $T_0=1$ (dashed lines).}
\label{Fig.Four}
\end{figure}

We start the discussion with the determination of the time $t_1$ defined by 
$E_{T_1}(t_1)=E_{T_2}^{\rm eq}$. 
The dependence on the temperature-difference $T_2-T_1$ of $t_1$ is shown in Fig.\ref{Fig.Four} 
starting from $T_0=\infty$ and $T_0=1$ for different $T_2$.
The most important feature that becomes evident immediately is the non-monotonic behavior of $t_1$ as a function of $T_2-T_1$ which to the best of our knowledge has not been observed previously in theoretical or experimental investigation of the Kovacs effect. 
This non-monotonic behavior might be rationalized if one compares $t_1$ to the times required to reach thermal equilibrium at $T_1$, $\t_{\rm eq}$. For the temperature range considered, 
$\t_{\rm eq}$ changes by more than four orders of magnitude, cf. Fig.\ref{Fig.Two}.
Therefore, we interprete the non-monotonic behavior in the following way:
\begin{itemize}
\vspace{-0.15cm}
\item[]$T_1\ll T_2$:
$t_1$ is long, because $\t_{\rm eq}(T_1)$ is long. 
Thus, it takes a long time to reach  $E_{T_1}(t_1)=E_{T_2}^{\rm eq}$.
\item[]$T_1\to T_2$:
Here, $E_{T_1}^{\rm eq}$ and $E_{T_2}^{\rm eq}$ take on very similar values.
Accordingly, $E_{T_2}(t)$ has to relax almost to equilibrium in order to reach 
$E_{T_2}^{\rm eq}$. In the limit $T_1\to T_2$ $t_1$ diverges.
\end{itemize} 
Of course, the temperature at which the minimum value of $t_1$ is observed depends on the stretching of the relaxation function and also on the absolute relaxation strength 
($E_{T_1}^{\rm eq}-E_{T_2}^{\rm eq}$ in our case).
Therefore, it is not easy to predict how large a temperature range has to be considered in a real experiment.
We speculate that the reason why this anomalous behavior has not been observed up to now is that one has to use quite large temperature differences. This, however, should not provide a serious problem in possible experimental realizations or computer simulations.

\begin{figure}[!h]
\vspace{-0.2cm}
\centering
\includegraphics[width=7.5cm]{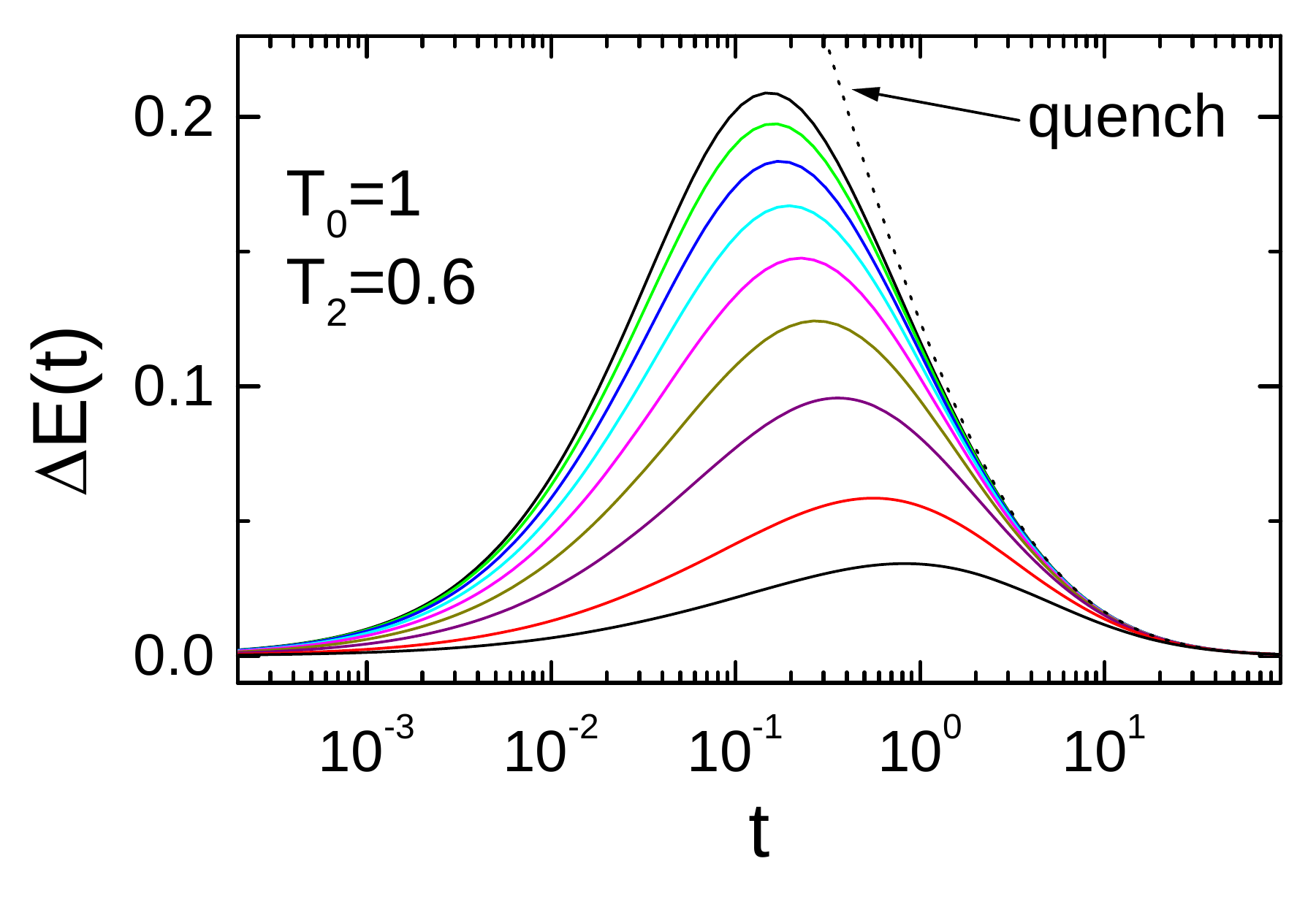}
\vspace{-0.3cm}
\caption{'Kovacs-hump' vs. $t$; 
$T_0=1\to T_2=0.6$ and various $T_1$, $T_1=0.2, 0.25, 0.3, 0.35, 0.4, 0.45, 0.5, 0.55, 0.575$ from top to bottom.}
\label{Fig.Five}
\end{figure}

Next, let us consider the Kovacs-hump $\D E(t)$ as defined in eq.(\ref{DelE.def}).
This quantity evolves as expected and as has been found in many earlier investigations.
For $T_0=1\to T_2=0.6$ it is plotted in Fig.\ref{Fig.Five}.
It is evident that for increasing $T_1$ the hump shifts towards longer times and becomes less intense. 
Therefore, the maximum
\be\label{DEK.def}
\D E_K=\left.\D E(t)\right|_{max}
\ee
increases as a function of the temperature difference $(T_2-T_1)$ and the time of the occurence of the maximum, 
\be\label{tauK.def}
\t_K:\quad \D E(t=\t_K)=\D E_K
\ee
decreases.

\begin{figure}[!h]
\vspace{-0.3cm}
\centering
\includegraphics[width=7.5cm]{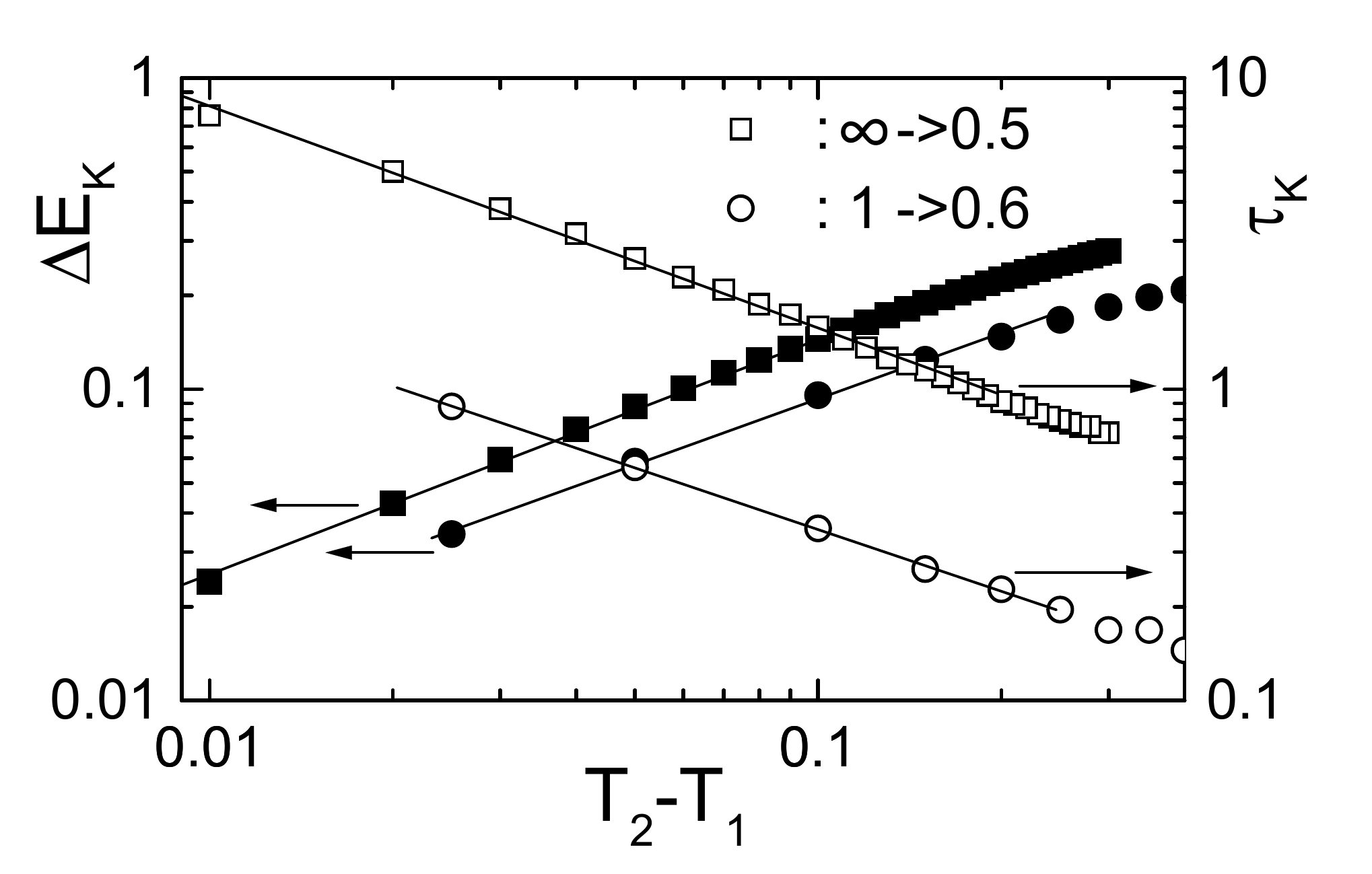}
\vspace{-0.3cm}
\caption{$\D E_K$ and $\t_K$ as a function of $(T_2-T_1)$; 
the lines are power laws $\D E_K\simeq a\cdot(T_2-T_1)$ and 
$\t_K\simeq b/(T_2-T_1)$.
($\infty\to 0.5$: $a=0.75$, $b=0.71$ and 
$1\to 0.5$: $a=0.70$, $b=0.66$.
}
\label{Fig.Six}
\end{figure}

Fig.\ref{Fig.Six} shows $\D E_K$ and $\t_K$ plotted versus $(T_2-T_1)$. 
It is obvious how $\D E_K$ increases and $\t_K$ decreases almost power-law like as a function of the temperature difference as also observed for other models\cite{Cugliandolo:2004p2938}.

\begin{figure}[!h]
\vspace{-0.2cm}
\centering
\includegraphics[width=7.5cm]{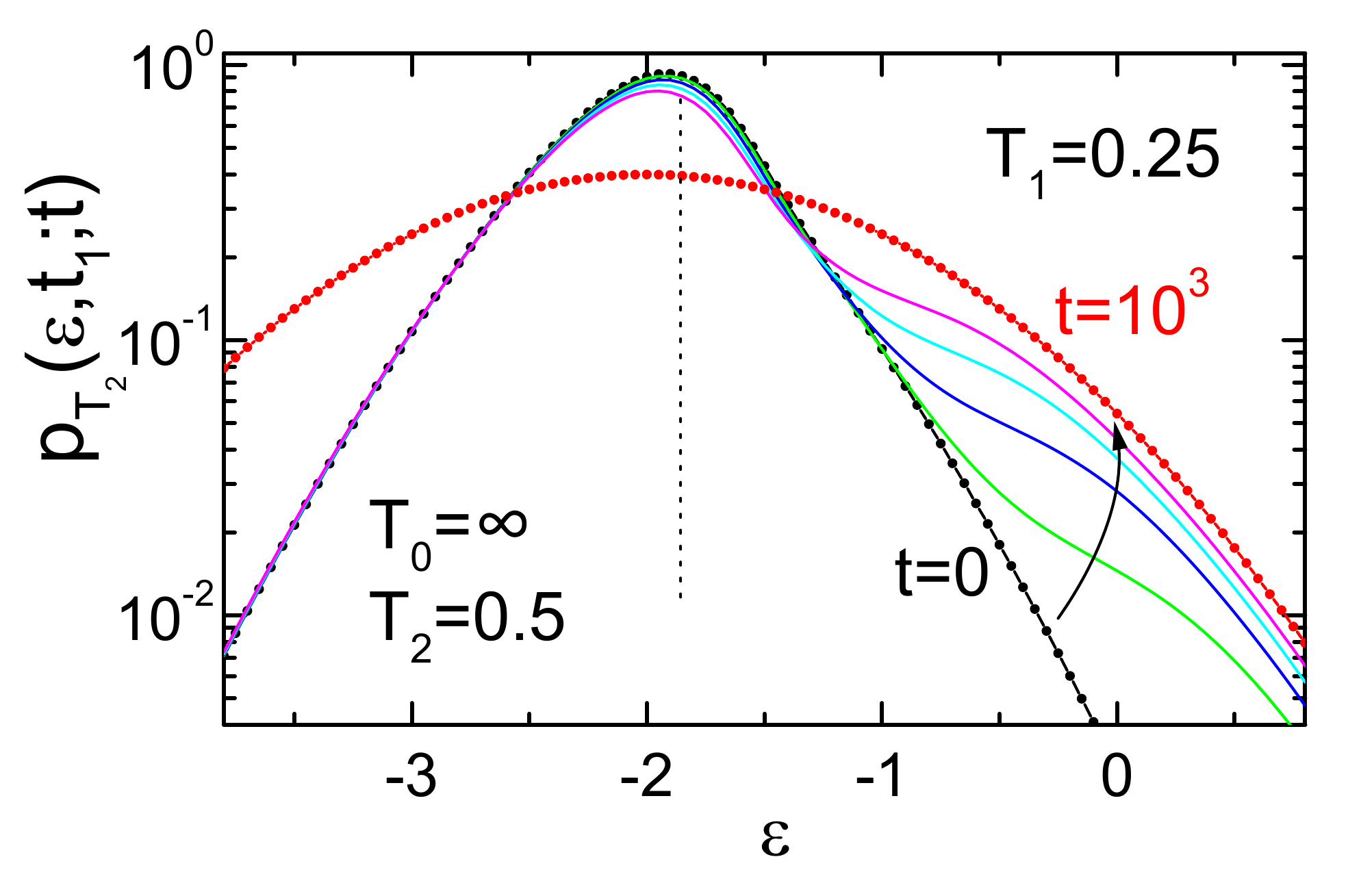}
\vspace{-0.3cm}
\caption{Populations for the Kovacs protocol after the temperature jump from $T_1$ to the final temperature $T_2=0.5$ after $t=0,0.01,0.03,0.06,0.1, 10^3$ from bottom to top 
as indicated by the arrow.
The dotted vertical line indicates the demarcation energy $\e_D(t_1)=-T_1\ln{(\k_\infty t_1)}$ and the symbols are for $t=0$ (black) and $t=10^3$ (red).}
\label{Fig.Seven}
\end{figure}

In order to see where the Kovacs-hump stems from in the present model, let us consider the populations in some more detail.
In Fig.\ref{Fig.Seven}, we plot the populations $p_{T_2}(\e,t_1;t)$ defined implicitly in eq.(\ref{E.Kovacs}). 
After a quench from $T_0=\infty$ to $T_1=0.25$ the temperature is changed to $T_2=0.5$ at the time $t_1$ needed for $E_{T_1}(t_1)=E_{T_2}^{\rm eq}$ given by $E_{T_2}^{\rm eq}=-2\s^2$.
For $t=0$, the population $p_{T_2}(\e,t_1;t=0)=p_{T_1}(\e,t_1)$ is just the same one as immediately before the temperature change from $T_1$ to $T_2$.
This distribution (the narrowest one in Fig.\ref{Fig.Seven}) is slightly asymmetric and quite narrow compared to the equilibrium distribution, which is in complete harmony with the results for a quench as discussed in ref.\cite{NicChris:2010}.
Note, that by definition of $t_1$ the distribution $p_{T_1}(\e,t_1)$ yields $E_{T_2}^{\rm eq}=-2\s^2$ for the mean energy.

For finite times $t$ a strong tailing of the distributions $p_{T_2}(\e,t_1;t)$ at higher energies is observed.
Conversely, for low energies all curves but the one for the longest time superimpose indicating that 
relaxation towards the equilibrium distribution takes place only after very long times.
This means that essentially all states with energies smaller than the so-called demarcation energy 
$\e_D(t_1)=-T_1\ln{(\k_\infty t_1)}$ (shown as the dotted vertical line) are frozen for a long 
time\cite{DYRE:1995p2812, DYRE:1987p2799, Arkhipov1979}.
For intermediate times (all but the longest) the distributions qualitatively show the appearance of a superposition of two distinct distributions, one essentially given by $p_{T_1}(\e,t_1)$ and another  one with more weight at higher energies and increasing intensity for longer times.
We mention that this behavior is reminiscent of that found in the exponential trap 
model\cite{Bertin:2003p2817}.

From the behavior observed in Fig.\ref{Fig.Seven}, the emergence of a Kovacs-hump becomes clear.
The mean value of the energy, $E_{T_2}(t_1+t)$, will first increase and only after the low-energy states start to relax, it will decrease again and finally yield $\D E(t)=0$.

The above discussion indicates that the population can be given as a superposition of two distributions.
This means that the Kovacs hump can approximately be written as a difference of two relaxation functions. 
In the following, we will show that for the trap model such an approximation can be deduced from the master equation,
${\dot p}_T(\e,t)= -\k_T(\e)p_T(\e,t)+\rho(\e)\!\int\!d\e'\k_T(\e')p_T(\e',t)$, which is to be complemented by an initial condition $p_T(\e,0)$.
In the second term, the gain-term, of the master equation, we approximate $p_T(\e',t)$ by its long-time limit, the equilibrium distribution $p_T^{\rm eq}(\e')$. 
Furthermore, we utilize detailed balance in the form 
$\int\!d\e'\rho(\e)\k_T(\e')p_T^{\rm eq}(\e')=\int\!d\e'\rho(\e')\k_T(\e)p_T^{\rm eq}(\e)
=\k_T(\e)p_T^{\rm eq}(\e)$ to obtain the approximative equation
${\dot p}_T(\e,t)= -\k_T(\e)p_T(\e,t)+\k_T(\e)p_T^{\rm eq}(\e)$.
The solution of this inhomogeneous equation poses no problem\cite{NicChris:2010} and for the Kovacs protocol, i.e. the initial condition $p_T(\e,0)=p_{T_1}(\e,t_1)$, one finds
\be\label{p.app.T2}
p^{\rm ap.}_{T_2}(\e,t_1;t)=N(t,t_1)^{-1}\left[p^{\rm eq}_{T_2}(\e)
			+\left\{p_{T_1}(\e,t_1)-p^{\rm eq}_{T_2}(\e)\right\}e^{-\k_{T_2}(\e)t}\right]
\ee
where 
$N(t,t_1)=1+\int\!d\e\left\{p_{T_1}(\e,t_1)-p^{\rm eq}_{T_2}(\e)\right\}\!e^{-\k_{T_2}(\e)t}$ is a normalization constant.
The dependence of $N(t,t_1)$ on $t_1$ can usually be neglected for a large part as can be seen from 
Fig.\ref{Fig.Seven}.
For small energies one has $p_{T_1}(\e,t_1)\ll p^{\rm eq}_{T_2}(\e)$ and thus
$N(t,t_1)\simeq N(t)=1-\int\!d\e p^{\rm eq}_{T_2}(\e)e^{-\k_{T_2}(\e)t}$.
For $\e>\e_D(t_1)$, the relaxation is fast meaning that $e^{-\k_{T_2}(\e)t}\simeq0$ and consequently
$N(t,t_1)\simeq1$.
Numerically, neglecting the dependence of $N(t,t_1)$ on $t_1$ introduces a maximum error (for times 
$t\sim t_1$) on the order of 10\%.
For other temperatures $T_1$ and $T_2$, the neglect of the $t_1$-dependence of $N(t,t_1)$ introduces an error of similar magnitude.

For the distribution of the populations following a quench, the approximative solution has been compared with the exact numerical solution of the master equation in ref.\cite{NicChris:2010} and good agreement has been found.
We note, that a similar agreement is observed for the $p_{T_2}(\e,t_1;t)$.
Eq.(\ref{p.app.T2}) shows that the Kovacs hump approximately will be given as the difference of two functions.
With the further assumption $N(t,t_1)\simeq N(t)$, one finds that one function depends on $t_1$ and $t$ and the other one only on $t$,
\be\label{DE.approx}
\D E(t)\simeq\phi(t_1;t)-\psi(t)
\ee
with 
$\phi(t_1;t)=N(t)^{-1}\int\!d\e\!\cdot\!\e\!\cdot\! p_{T_1}(\e,t_1)e^{-\k_{T_2}(\e)t}$ and
$\psi(t)=N(t)^{-1}\int\!d\e\!\cdot\!\e\!\cdot\! p^{\rm eq}_{T_2}(\e)e^{-\k_{T_2}(\e)t}$.
A behavior of the hump similar to eq.(\ref{DE.approx}) has been proposed earlier in ref.\cite{Bertin:2003p2817} (eq.(27)) and it is also similar to what is obtained in linear response theory, cf. below.
\subsubsection*{2. 'Inverse' Kovacs experiment}
\begin{figure}[!h]
\centering
\includegraphics[width=6cm]{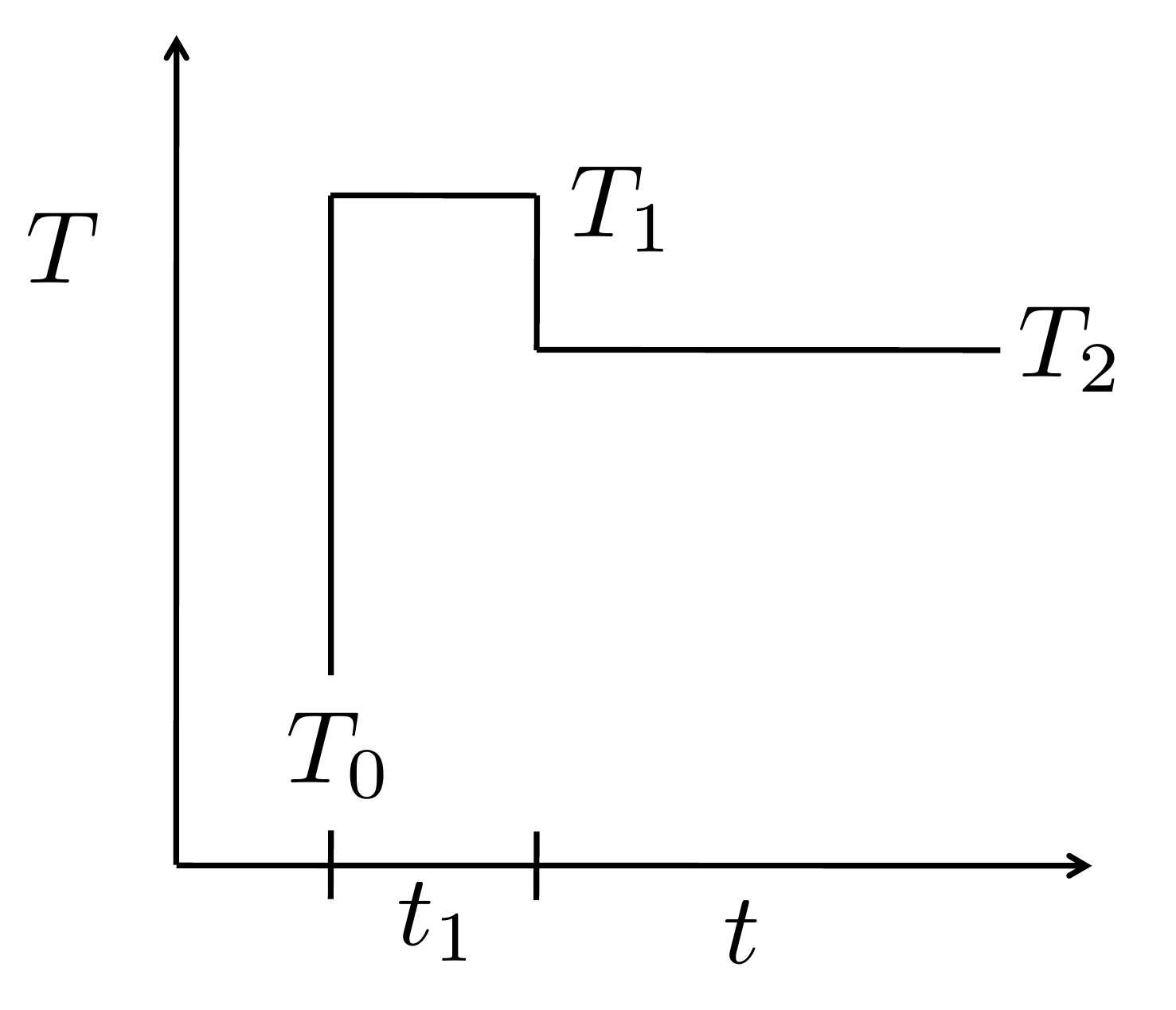}
\vspace{-0.5cm}
\caption{Sketch of the protocol for the Kovacs effect starting from a low temperature.}
\label{Fig.Eight}
\end{figure}

\begin{figure}[!h]
\vspace{-0.2cm}
\centering
\includegraphics[width=7cm]{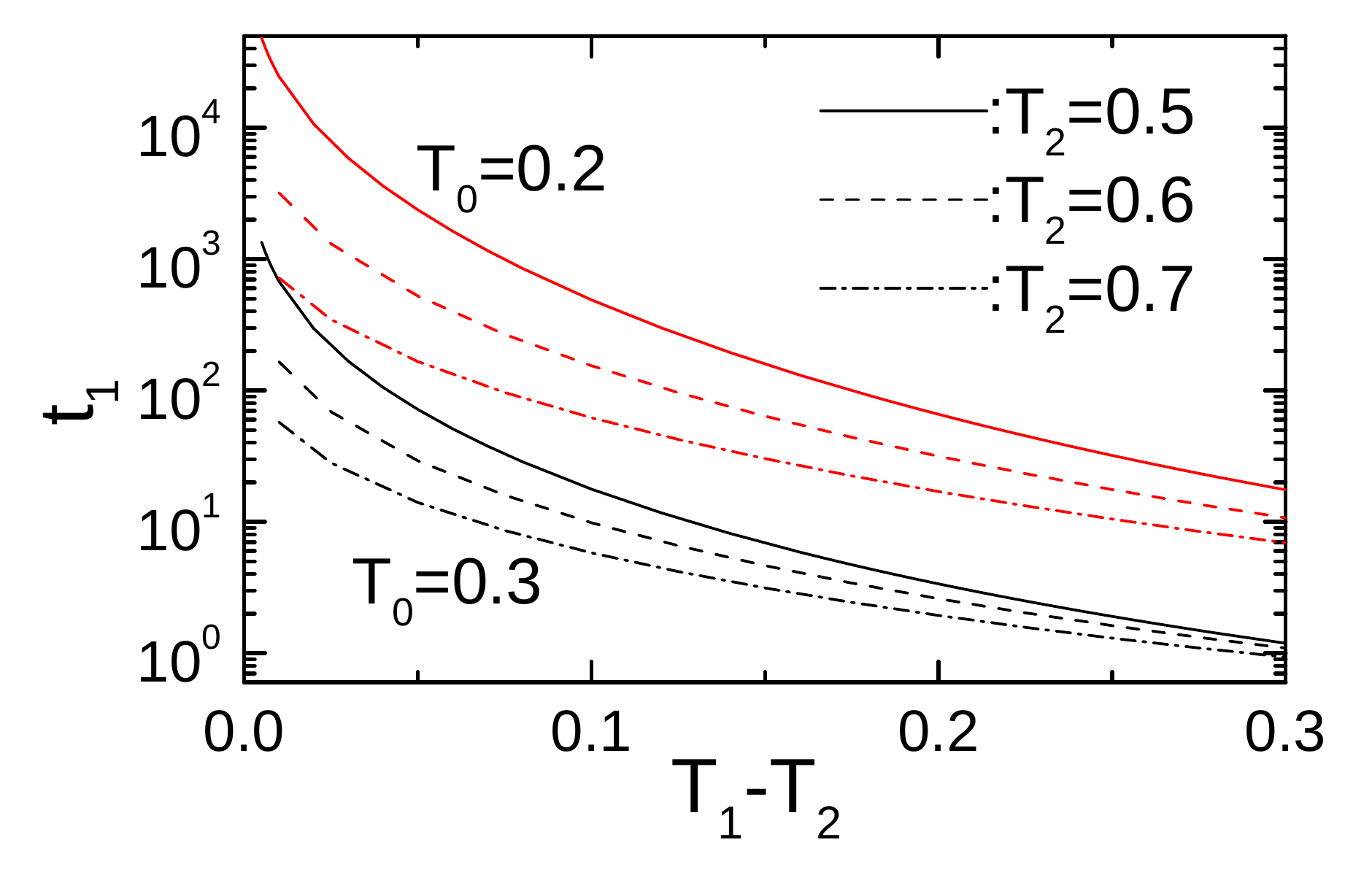}
\vspace{-0.5cm}
\caption{$t_1$ versus $(T_1-T_2)$ ($T_2=0.5, 0.6, 0.7$) for the Kovacs protocol starting from $T_0=0.2$ (upper red lines) and $T_0=0.3$ (lower black lines).}
\label{Fig.Nine}
\end{figure}

Next, let us consider the Kovacs effect starting from a low temperature instead from a high temperature.
The protocol then assumes the following form, see Fig.\ref{Fig.Eight}.
%
%\vspace{-0.0cm}
As in the case of a quench from $T_0>T_1$, $T_2$ (cf. Figs.\ref{Fig.Three},\ref{Fig.Four}), we first determine the time $t_1$ at which $E_{T_1}(t_1)=E_{T_2}^{\rm eq}$ and the results are shown in Fig.\ref{Fig.Nine}.
In contrast to the behavior for the standard Kovacs protocol, in this case $t_1$ changes monotonously as a function of temperature difference.
This is because $T_1$ is the highest temperature and therefore the relaxation time at $T_1$ is the shortest.
Only for $T_1\to T_2$ it takes a long time to reach $E_{T_2}^{\rm eq}$ because the energy has to relax almost to its equilibrium value.

\begin{figure}[!h]
\vspace{-0.2cm}
\centering
\includegraphics[width=7cm]{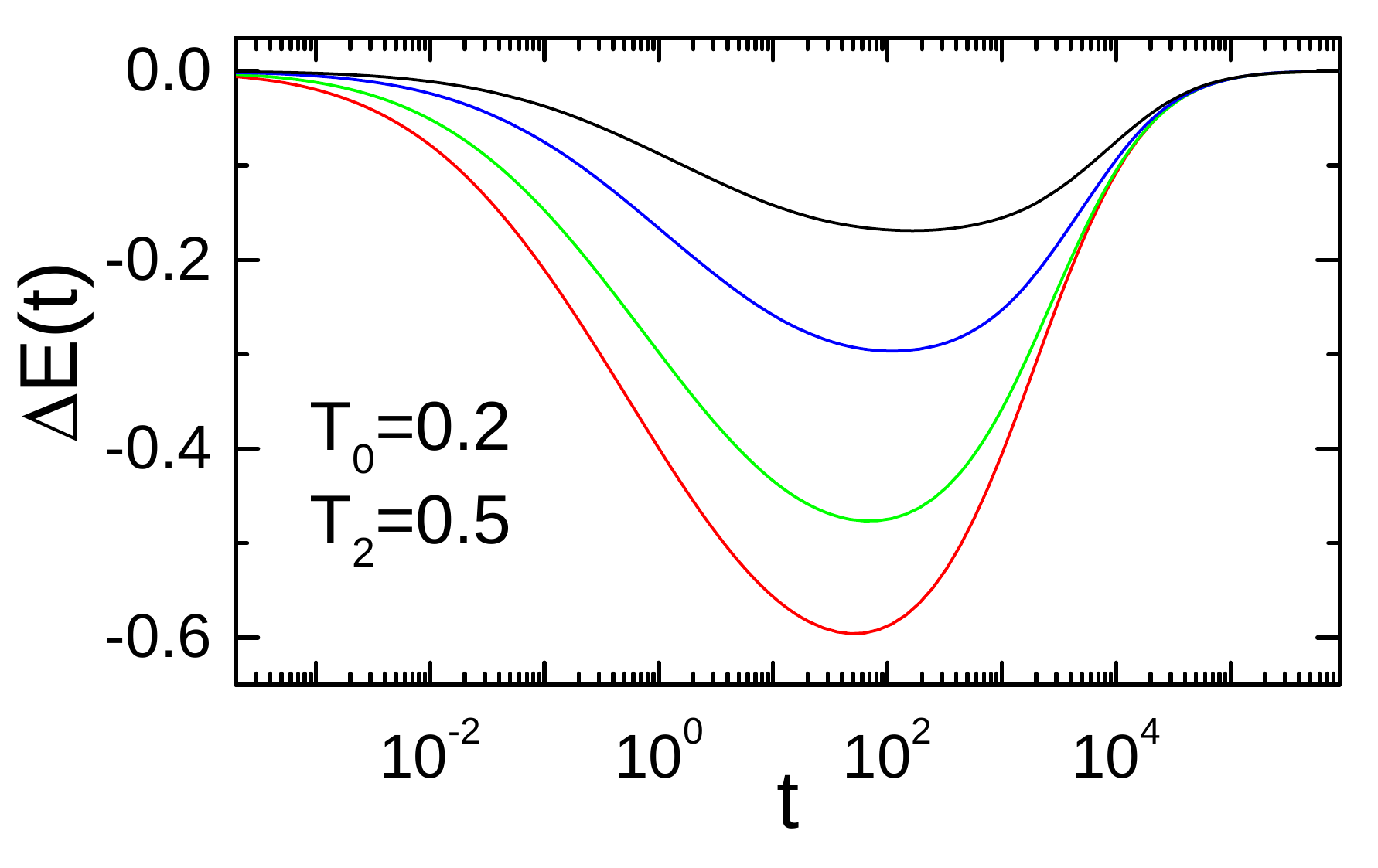}
\vspace{-0.3cm}
\caption{Kovacs-hump for $T_0=0.2\s$ and $T_2=0.5\s$ and various values of $T_1$($>T_2$),
$T_1=0.55, 0.6, 0.7, 0.8$ from top to bottom.}
\label{Fig.Ten}
\end{figure}

An example for the Kovacs-hump is shown in Fig.\ref{Fig.Ten} for $T_0=0.2$ and $T_2=0.5$.
The behavior is - apart from the sign - similar to what is observed in Fig.\ref{Fig.Five} for a quench from a high temperature.
The minimum $\D E_K$ increases with increasing $|T_2-T_1|$ and $\t_K$ decreases.

\begin{figure}[!h]
\vspace{-0.2cm}
\centering
\includegraphics[width=7cm]{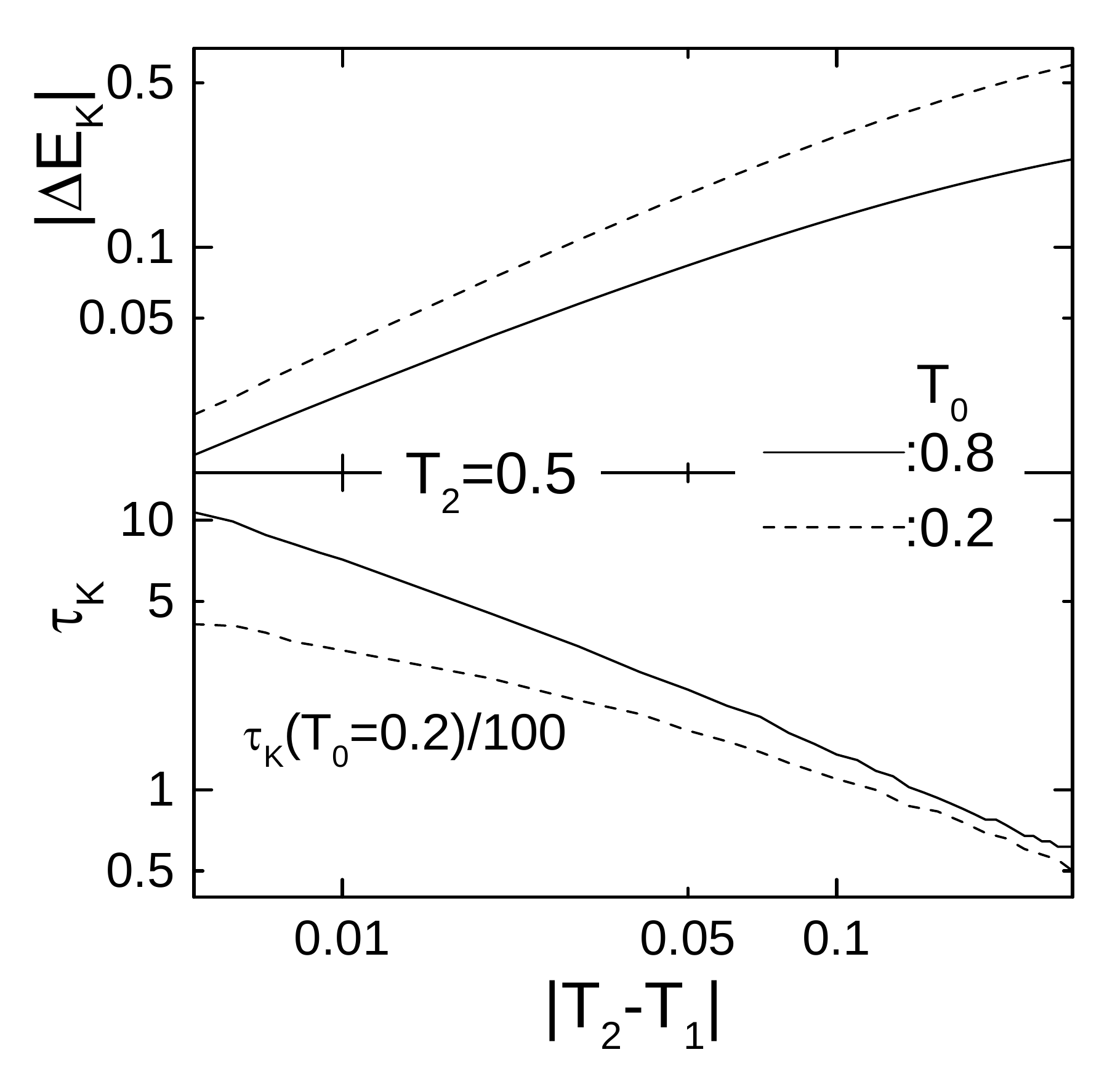}
\vspace{-0.50cm}
\caption{$|\D E_K|$ (upper panel) and $\t_K$ (lower panel) vs. $|T_2-T_1|$ for $T_0=0.8$ and $T_0=0.2$.}
\label{Fig.Eleven}
\end{figure}

For a direct comparison with the results of the standard protocol, in Fig.\ref{Fig.Eleven} we plot 
$\D E_K$ and $\t_K$ versus $|T_2-T_1|$ for $T_0=0.8$ and for $T_0=0.2$. 
The most important feature of this plot is the similarity in the characteristics of the Kovacs hump for high and low initial temperature $T_0$.
In both cases, one has power-law behavior of $\D E_K$ and $\t_K$ as a function of $|T_2-T_1|$ for small temperature differences.
However, also the well known asymmetry of the relaxation\cite{Angell:2000p3651} becomes obvious via the much longer time $\t_K$ for the jump from a low temperature.

\begin{figure}[!h]
\vspace{-0.2cm}
\centering
\includegraphics[width=7cm]{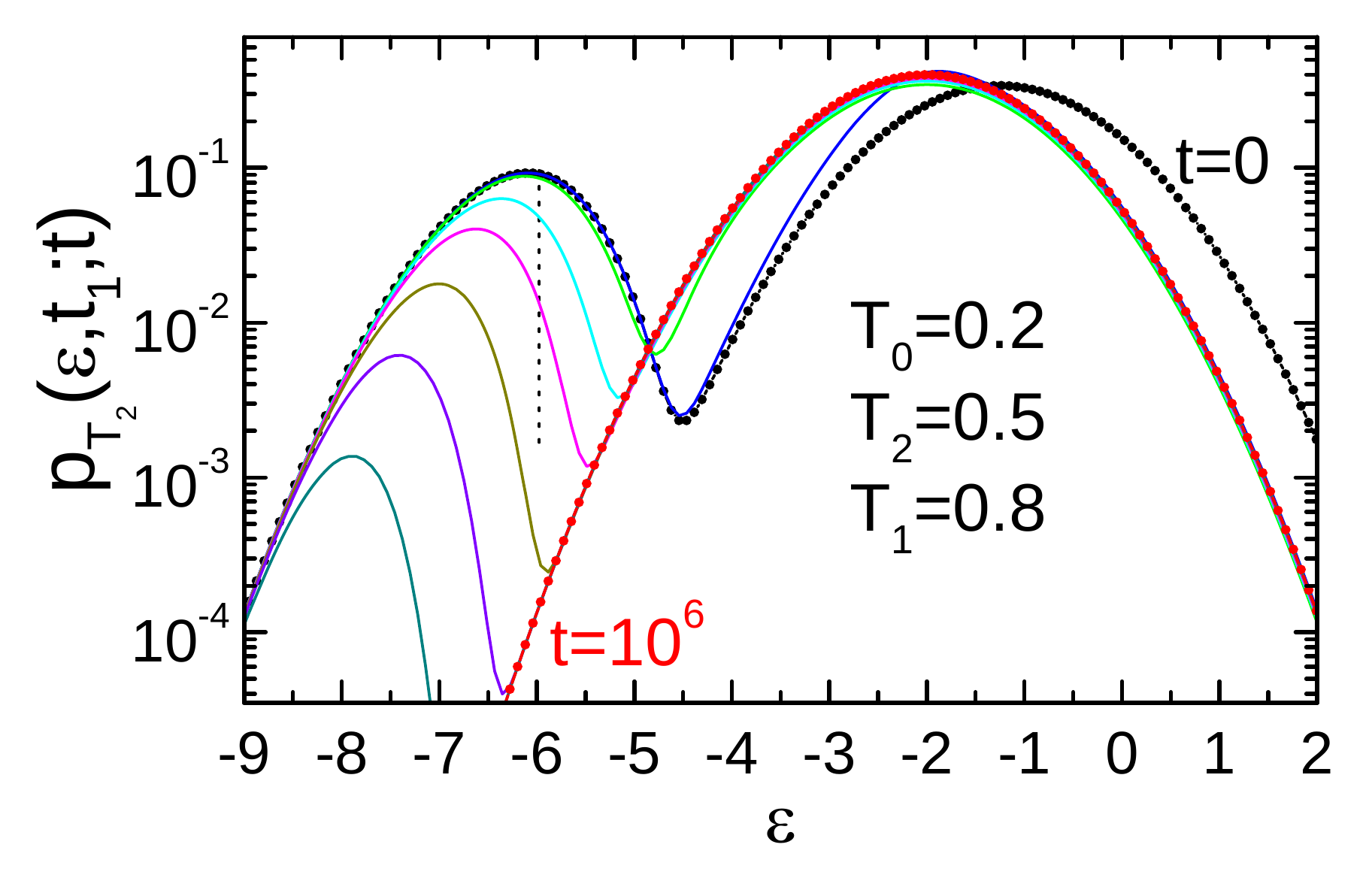}
\vspace{-0.3cm}
\caption{Populations $p_{T_2}(\e,t_1;t)$ following the inverse Kovacs protocol for various times 
$t=0,10^2,10^3,3\cdot\!10^3,10^4,3\cdot\!10^4,10^5,10^6$ (from top to bottom).
The intensity of the low-energy peak decreases with increasing time, indicating that states with
$\e<\e_D$ (dotted vertical line) relax.
The symbols are for $t=0$ (black) and $t=10^6$ (red).
}
\label{Fig.Twelve} 
\end{figure}

The fact that there is no qualitative difference between the two protocols seems somewhat puzzling on first sight.
As mentioned already above, in case of an up-jump, the distribution of energies, $p_T(\e,t)$, shows a two-peak structure for intermediate times.
The same behavior is observed when considering $p_{T_2}(\e,t_1;t)$ for various times as shown in 
Fig.\ref{Fig.Twelve}.
From this plot one can see how the states with energies smaller than the demarcation energy $\e_D(t_1)$ relax as time progresses.
The reason for the fact that this two-peak structure of the populations is not reflected in the shape and position of the Kovacs hump lies in the inherent averaging performed when the energy is calculated according to eq.(\ref{E.n.t}).
As in the 'standard' protocol, the Kovacs hump will approximately be given by a difference of two relaxation functions, cf. eq.(\ref{DE.approx}).

The expression given in eq.(\ref{DE.approx}) along with the inspection of the populations for the two protocols can be used to give a natural explanation for the sign of the Kovacs hump.
For short times, the functions $\phi$ and $\psi$ can be written as
$\phi(t_1;t)\simeq E_{T_2}^{\rm eq}-t\int\!d\e\!\cdot\!\e\!\cdot\!p_{T_1}(\e,t_1)\k_{T_2}(\e)$ and
$\psi(t)\simeq E_{T_2}^{\rm eq}-t\int\!d\e\!\cdot\!\e\!\cdot\!p^{\rm eq}_{T_2}(\e)\k_{T_2}(\e)$.
Now, assuming a not too low temperature $T_2$, one can further approximate
$\k_{T_2}(\e)\simeq\k_\infty e^{\b_2E_{T_2}^{\rm eq}}(1+\b_2(\e-E_{T_2}^{\rm eq})+\cdots)$.
With this, one finds
\[
\D E(t)\simeq\b_2(\k_\infty t)e^{\b_2E_{T_2}^{\rm eq}}
\left(V_\infty-V_0\right)
\]
where the variances are defined as 
$V_\infty=\int\!d\e\!\cdot\!(\e-E_{T_2}^{\rm eq})^2\!\cdot\!p^{\rm eq}_{T_2}(\e)$ and 
$V_0=\int\!d\e\!\cdot\!(\e-E_{T_2}^{\rm eq})^2\!\cdot\!p_{T_1}(\e,t_1)$
(remember that $p_{T_1}(\e,t_1)\equiv p_{T_2}(\e,t_1;t=0$)).
Thus, a glance at Fig.\ref{Fig.Seven} and Fig.\ref{Fig.Twelve} shows that for the standard protocol one has $\D E(t)\geq0$ and for the inverse protocol $\D E(t)\leq0$ holds.

Even though the Kovacs effect reflects the complexity of the relaxation properties of the system in some sense, its overall behavior does not depend very sensitive on the underlying distribution of relaxation times.
This finding might also explain why different models yield very similar results concerning the Kovacs effect.
According to what has been discussed above the appearence of the Kovacs hump is a consequence of the fact that some energies have already relaxed at the time $t_1$ of the jump from $T_1$ to $T_2$ while others are still out of equilibrium.
From Fig.\ref{Fig.Seven} and Fig.\ref{Fig.Twelve} and the discussion of the role of the demarcation energy it is clear that in both cases one might be able to model the Kovacs hump as the difference of two functions, cf. eq.(\ref{DE.approx}).
This results in a situation that is very reminiscent of what is observed in linear response theory\cite{Boehmer:1995p4162} which we will discuss in the following.
\subsubsection*{3. Linear response theory}
\begin{figure}[!h]
\vspace{-0.2cm}
\centering
\includegraphics[width=7cm]{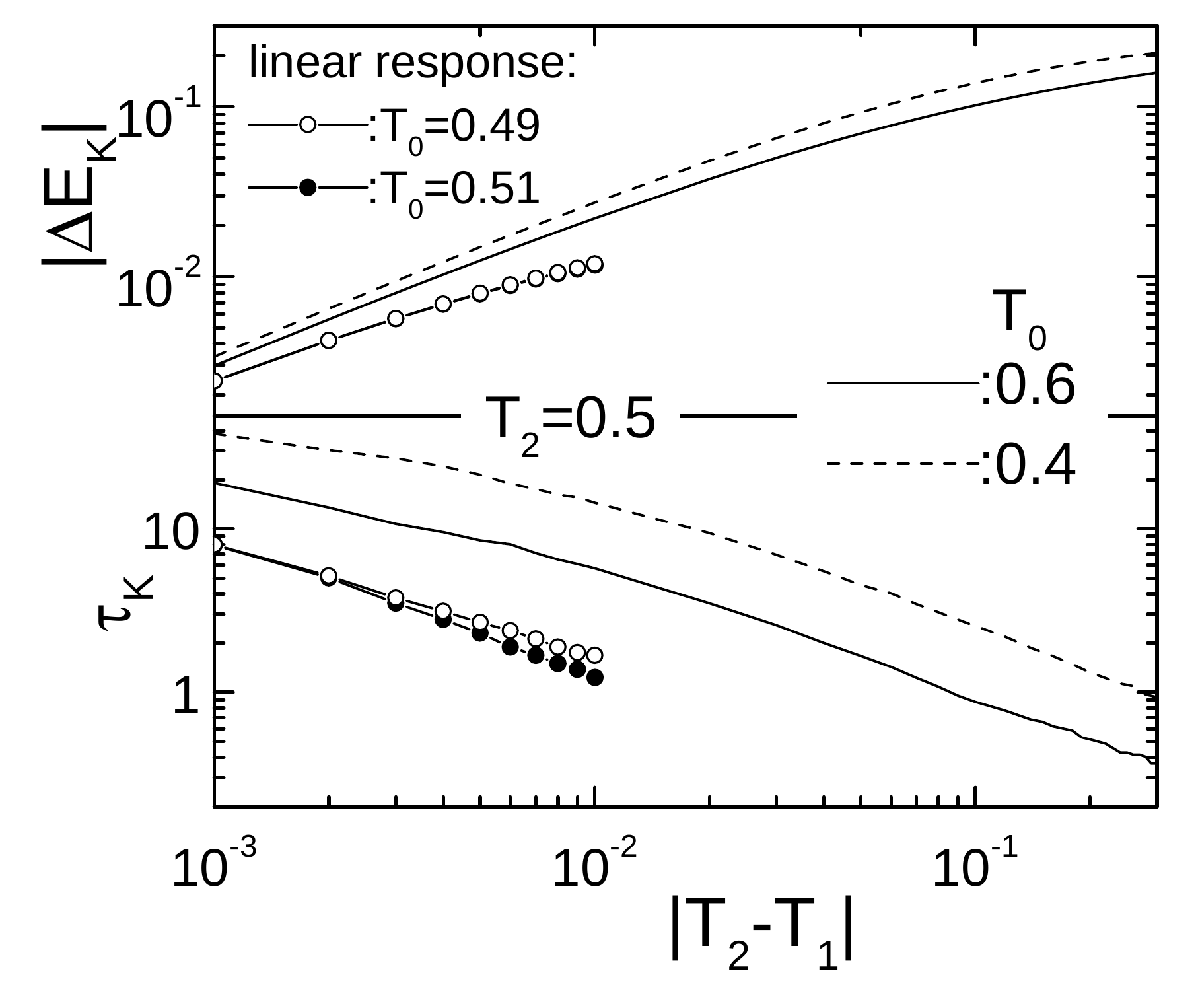}
\vspace{-0.50cm}
\caption{$\D E_K$, $\t_K$ vs. $|T_2-T_1|$ for different initial temperatures $T_0$}
\label{Fig.Thirteen}
\end{figure}

Using the superposition principle of linear response theory, one can calculate the response of the Kovacs protocol for $|T_0-T_1|\to 0$ and $|T_2-T_1|\to 0$ in the same way as in the calculation for a single temperature step, cf. eq.(\ref{CN.eq}), with the result (cf. ref.\cite{Boehmer:1995p4162}):
\be\label{Kovacs.eq.linresp}
E_{T_2}(t_1+t)=E_{T_2}^{\rm eq}+\left[E_{T_0}^{\rm eq}-E_{T_1}^{\rm eq}\right]C_{\rm eq}(t_1+t)
-\left[E_{T_2}^{\rm eq}-E_{T_1}^{\rm eq}\right]C_{\rm eq}(t)
\ee
where $C_{\rm eq}(t)$ is evaluated at $T_0$ and again $\D E(t)=E_{T_2}(t_1+t)-E_{T_2}^{\rm eq}$.
As mentioned already in the Introduction, one is concerned with the difference of two relaxation functions evaluated at the same temperature but at different times.
Of course, linear response theory only holds for very small temperature jumps.
Instead of showing results for the Kovacs hump here, we note that these qualitatively are very similar independent of the magnitude of the temperature differences.
Only the quantitative features depend on the temperature regime considered.
If one starts from a low temperature instead, the results are very similar. 
In Fig.\ref{Fig.Thirteen} we show the results for the peak, $|\D E_K|$ (eq.(\ref{DEK.def})), and the time of its appearance, $\t_K$ (eq.(\ref{tauK.def})), as a function of the temperature difference for some very small and some large temperature differences.
From this plot it becomes apparent that the Kovacs effect does not change its qualitative behavior when leaving the linear response regime.
The small differences in the values for $\t_K$ for the positive and the negative difference show the limitations of the linear response regime and thus the onset of the asymmetry.

Eq.(\ref{Kovacs.eq.linresp}) directly shows that a Kovacs hump will only be observed if the energy relaxation function $C_{\rm eq}(t)$ decays in a non-exponential fashion, as discussed in 
ref.\cite{Boehmer:1995p4162}. 
Also in the nonlinear regime we find that the Kovacs hump can approximately be described as the difference between two relaxation functions, albeit they are not the same.
Therefore, here a Kovacs hump can be observed also for exponentially decaying functions, if the relaxation times of $\phi$ and $\psi$ are different.
However, for the trap model, all relevant functions decay non-exponentially.
A notable exception from this behavior seems to be provided by the relaxation function in the p-spin model if one chooses $p=3$, as is done in the quoted model calculations\cite{Cugliandolo:2004p2938}. 
In this case, the long-term relaxation in equilibrium is given by a single exponential function
($C_{\rm eq}(t)\sim e^{-t}$), but the relevant relaxation functions in the nonlinear regime are non-exponential.
However, the relevance of such a behavior for the interpretation of the Kovacs effect in supercooled liquids is not obvious.
Another model that shows exponential relaxation is the quoted two-state model\cite{Aquino:2008}.
Here, the non-equilibrium situation is imposed by quenches with a finite cooling rate, a situation that we do not consider in the present paper.
We expect that both models do not show a Kovacs effect in the linear response regime.

If we relax the requirement that the temperature jump $T_0\to T_1$ be small, but consider only small $|T_2-T_1|$, the Kovacs protocol can be viewed as a linear response experiment performed on an aging system.
In this case, one can use the formalism described in detail in ref.\cite{Diezemann:2005p4163}.
One just has to recognize that now the temperature jump from $T_1$ to $T_2$ plays the role of an applied external field, $H=(T_2-T_1)/T_1$ (and the parameters determining the coupling of the field are to be chosen as $\mu=1$ and $\g=0$). 
The response is then determined by the energy autocorrelation function, where now this function is evaluated under the non-equilibrium situation. 
\section*{III. Conclusions}
In the present paper we have investigated the memory effect or Kovacs effect in a simple model for glassy relaxation, the Gaussian trap model. 
This model is interesting because the system can reach thermal equilibrium at all temperatures and it captures some aspects of the relaxation observed in real glass-forming liquids and computer models.

The relaxation of the system following the Kovacs protocol shows a behavior that is similar to what has been found in earlier investigations on other models and also experimentally.
However, we find one distinctive feature in our model calculations that has to our knowledge not been observed previously.
The temperature dependence of the time $t_1$, which is defined by the condition 
$E_{T_1}(t_1)=E_{T_2}^{\rm eq}$, shows a non-monotonic behavior and it first decreases, goes through a minimum and then increases again as a function of the temperature difference $|T_2-T_1|$.
We suggest that $|T_2-T_1|$ has to be rather large in order to observe this effect. 
Thus, we would propose to perform experiments for larger temperature differences.
However, a quantitative estimate of the required temperature differences is hampered by the fact that this depends on the relaxation strength and thus on the variable considered.

For an up-jump and then an 'inverse' Kovacs protocol we find a behavior that is very similar to that following a quench, the main differences being a different sign of the Kovacs hump and a monotonic behavior of $t_1$. 
The two-peak structure of the energy distribution does not affect the Kovacs hump significantly due the inherent averaging in the calculation of the mean energy.

In both cases the different relaxation behavior for states with energies smaller and larger than the demarcation energy can be used to rationalize the origin of the Kovacs hump.
Low energy states are still frozen at $t_1$, the time of the temperature jump to $T_2$, while large energy states respond quickly. 
From this fact it follows that the Kovacs hump can approximately be described as a difference of two relaxation functions, $\D E(t)\simeq \phi(t_1,t)-\psi(t)$.
The sign of the hump can be rationalized by an explicit short-time expansion.
The concept of the demarcation energy is quite general and it can be used to understand many features observed in the aging dynamics of glass-forming liquids including the main features of the Kovacs effect.
We interprete the fact that many models exhibiting a broad distribution of relaxation times qualitatively show a similar behavior as stemming from the distinction of slowly and fast relaxing entities. 
It is possible to relate the mobility of the entity to its potential energy. 
This establishes a direct connection between the microscopic system and the energy-related trap 
model\cite{Heuer:2008p3506}.
The similarities found in different models can be understood as stemming from the fact that the averaging inherent in the evaluation of bulk properties is not very sensitive to the detailed shape of the underlying distributions. 
Consequently, the Kovacs protocol does not seem very well suited to distinguish among different models for dynamic heterogeneities in glass-forming liquids.
 
When the temperature differences all are small one can use linear response theory for the calculation showing that the Kovacs hump has its origin in the subtraction of the energy autocorrelation function evaluated at two different times with different weights.
We have shown that the qualitative behavior does not change for larger temperature differences. 
If only $|T_2-T_1|$ is small but $|T_0-T_1|$ is arbitrary, one also can use linear response theory but now one has to explicitly consider the non-equilibrium situation, resulting in a dependence of the energy autocorrelation function on two times.

We conclude with noting that the Kovacs hump for a given quantity can be qualitatively understood in terms of the linear response behavior. In the nonlinear regime mainly quantitative differences are found but we do not observe any qualitatively new features.
\section*{Acknowledgment}
We thank Roland B\"ohmer for drawing our attention to the behavior of the Kovacs hump in linear response theory. 
We thank him, Jeppe Dyre and Christian Rehwald for fruitful discussions.
A.H. acknowledges support from the CeNoS M\"unster.
\end{document}